\documentstyle[editedvolume]{crckapb}

\begin{opening}

\title{Spiral structure parameters\protect\\
       in the solar neighbourhood}

\author{D. Fern\'andez}
\author{X. Luri}
\author{F. Figueras}
\author{J. Torra}
\institute{Dpt. d'Astronomia i Meteorologia, Universitat de
           Barcelona, Av. Diagonal 647, E-08028 Barcelona, Spain}

\end{opening}

\runningtitle{Spiral structure parameters}

\begin{document}

\begin{abstract}

Two samples of O- and B-type stars and Cepheids with Hipparcos data have
been used to characterize galactic rotation and spiral arm kinematics in
the solar neighbourhood. An extensive set of simulations has been
performed in order to assess the capabilities of the method and its
sensitivity to sample errors and biases in the adopted parameters.

\end{abstract}

\section{Introduction}

Young stars are good tracers of the galactic spiral structure in the solar
neighbourhood, but, unfortunately, the observational data available up to
now have not allowed to reach conclusive results on the parameters
describing such structure. The accurate astrometric data provided by
Hipparcos (ESA 1997) allow us to re-examine the kinematic behaviour of
these stars and, even more, to evaluate the possibility to undertake this
study using more sophisticated kinematic models.

In a previous paper (Torra et al. 2000) we characterized the structure and
kinematics of the Gould Belt system, establishing its boundary to a
distance of 600 pc from the Sun. Beyond this distance, the kinematic
behaviour of young stars, in particular O-B stars and Cepheids, has to
reflect the smooth galactic potential, that is the galactic rotation and
the kinematics owing to the spiral structure. The study of the stellar
velocity field outside the Gould Belt is undertaken in the present paper,
considering both the current observational material -- its errors and the
sample's constraints -- and, through realistic simulations, the robustness
of the resolution process. In particular, for Cepheid stars, the influence
of the period-luminosity relation (PLR) on the derived kinematic
parameters is also evaluated.

\section{The working samples}

\subsection{Sample of O- and B-type stars}

The construction of this sample using Hipparcos data is fully described in
Fern\'andez (1998) and Torra et al. (2000). Due to the uncertainty on
trigonometric parallax at large distances, these were computed following
Crawford's (1978) photometric calibration. Preliminary results applying
the LM method (Luri et al. 1996) to Hipparcos data indicate this
calibration is good enough for our purposes (Jordi et al. 2000).
Considering only those stars with an heliocentric distance 0.6 $< R <$ 2
kpc, the sample with distances and proper motions contains 444 stars,
whereas the subsample with available radial velocities (Grenier 1997)
contains 304 stars.

\subsection{Sample of Cepheid stars}

The initial sample contains all the Hipparcos classical Cepheids.
Astrometric data were taken from this catalogue and radial velocities from
Pont et al. (1994, 1997). Two PLR have been used for distance computation:  
Luri's (1999) relation, which corresponds to the short cosmic distance
scale ($M_v = -1.08 - 2.72 \log P$), and Feast \& Catchpole (1997)
relation, which gives a long cosmic scale ($M_v = -1.41 - 2.81 \log P$).
In both cases, periods were obtained from Hipparcos and individual
reddenings from Fernie et al.'s (1995) compilation (continuous updating).
Up to $R =$ 4 kpc, the sample with known distances and proper motions
contains 203 stars and the subsample with known radial velocities contains
179 stars.

\section{Kinematic model}

Our kinematic model considers the systematic contributions of solar
motion, differential galactic rotation (up to second-order approximation)
and spiral arm kinematics (modeled in the frame of the Lin's theory). In
addition to the solar motion, we derived the first- and second-order terms
of the galactic rotation curve, the phase of the spiral structure at the
Sun's position, the amplitudes in the perturbation of the velocity in the
galactocentric and tangential directions and an adimensional parameter
which takes into account the difference in the velocity dispersion between
the solar-type stars and the considered stars. There are four free
parameters: the number of spiral arms of the galaxy ($m$), their pitch
angle ($i$), the galactocentric distance of the Sun ($\varpi_\odot$) and
the circular velocity at the Sun's position ($\Theta(\varpi_\odot)$). A
couple of values of these parameters were considered: $m =$ 2 or 4, $i =
-6^\circ$ or $-12^\circ$, $\varpi_\odot =$ 7.1 or 8.5 kpc and
$\Theta(\varpi_\odot) =$ 184 or 220 km s$^{-1}$. The kinematic parameters
were derived via weighted least squares fit, following an iterative
scheme. Three cases were studied, solving the equations considering only
radial velocities, only proper motions and the combined resolution (see
Fern\'andez et al. 2001).

\section{Test of robustness: simulations}

Numerical simulations were performed in order to quantitatively evaluate
the biases in the kinematic model parameters induced by, among others, the
observational errors, the incompleteness of the sample, the rejection of
high residual stars and the correlations among the parameters. Both O-B
star and Cepheid samples were simulated considering the same spatial
distribution than the real sample and a Schwarzschild distribution for the
residual velocities. Solar motion, galactic rotation and spiral arm
kinematics were introduced following our kinematic model. We assumed
gaussian errors in distance, radial velocity and proper motions.

Detailed conclusions of these simulations can be found in Fern\'andez et
al. (2001). As main results, we conclude that the combined solution
minimizes the biases induced by correlations. The observational errors in
proper motion, radial velocities and distance produce biases less than 0.5
km s$^{-1}$ in the solar motion components and in the galactic rotation
and spiral structure parameters. Due to the spatial distribution of the
available data, the best distance intervals were found to be 0.6 $< R <$ 2
kpc for O-B stars and $ R <$ 4 kpc for Cepheids.

\section{Results and discussion}

\subsection{Galactic rotation curve}

For the $A$ Oort constant, a discrepancy between the different solutions
has been found: $A^{\mathrm OB} \approx 13.6 \pm 0.7$ km s$^{-1}$
kpc$^{-1}$ for O-B stars and $A^{\mathrm Cep}_{\mathrm Luri} \approx 16.7
\pm 0.6$ km s$^{-1}$ kpc$^{-1}$, $A^{\mathrm Cep}_{\mathrm Feast} \approx
15.0 \pm 0.6$ km s$^{-1}$ kpc$^{-1}$ for Cepheids (Luri's 1999 and Feast
\& Catchpole's 1997 PLRs, respectively). This discrepancy has been
extensively found in the literature (e.g. Hanson 1987, Metzger et al.
1997, Feast \& Whitelock 1997 and Mishurov \& Zenina 1999 obtained values
between 11.3 and 18.8 km s$^{-1}$ kpc$^{-1}$) and, as pointed out by
Olling \& Merrifield (1998), it can be produced by the different spatial
distribution of the samples. In agreement with Feast et al. (1998), we
found a very small second-order term of the galactic rotation, compatible
with a linear rotation curve in the solar neighbourhood.

\subsection{Galactic spiral structure}

Using both O-B and Cepheid stars data, we found an angular rotation
velocity for the spiral pattern of $\Omega_{\mathrm{p}} \approx$ 30-35 km
s$^{-1}$ kpc$^{-1}$, incompatible with the classical value 13.5 km
s$^{-1}$ kpc$^{-1}$ proposed by Lin and collaborators in the later 60s.
Our value is in good agreement with the larger values recently published
by Amaral \& L\'epine (1997) and Mishurov et al. (1997, 1999). From this
value we can conclude that the Sun is outside -- though near -- the
corotation circle.

From the available data we are not able to decide on a galaxy with 2 or 4
spiral arms. Nowadays, this is an unsolved problem (see contradictory
results in Amaral \& L\'epine 1997, Mishurov \& Zenina 1999, Drimel 2000,
L\'epine et al. 2000). Nevertheless, good coherence is obtained to derive
the phase of the spiral structure: from both samples, we obtained values
in the rank [330$^\circ$,50$^\circ$], indicating the Sun is near the
potential minimum of an arm. The values of the velocity amplitudes due to
the spiral perturbation were found to be small, about 2 km s$^{-1}$.

\acknowledgements

This work has been suported by the CICYT under contracts ESP97-1803 and
ESP1999-1519-E.

\end{document}